\documentclass[sigconf]{acmart}

\AtBeginDocument{%
  \providecommand\BibTeX{{%
    \normalfont B\kern-0.5em{\scshape i\kern-0.25em b}\kern-0.8em\TeX}}}

\copyrightyear{2021}
\acmYear{2021}
\setcopyright{acmcopyright}
\acmConference[SIGIR '21]{Proceedings of the 44th International ACM SIGIR Conference on Research and Development in Information Retrieval}{July 11--15, 2021}{Virtual Event, Canada}
\acmBooktitle{Proceedings of the 44th International ACM SIGIR Conference on Research and Development in Information Retrieval (SIGIR '21), July 11--15, 2021, Virtual Event, Canada}
\acmPrice{15.00}
\acmDOI{10.1145/3404835.3462835}
\acmISBN{978-1-4503-8037-9/21/07}

\usepackage{multirow}
\usepackage[flushleft]{threeparttable}
\usepackage{subfigure}
\usepackage{enumitem}

\begin{document}

\title{UGRec: Modeling Directed and Undirected Relations for Recommendation}

\author{Xinxiao Zhao$^1$, Zhiyong Cheng$^2$, Lei Zhu$^3$, Jiecai Zheng$^4$, Xueqing Li$^1$}
\affiliation{\institution{$^1$School of Computer Science and Technology, Shandong University, China}}
\affiliation{\institution{$^2$Shandong Artificial Intelligence Institute, Qilu University of Technology (Shandong Academy of Sciences), China}}
\affiliation{\institution{$^3$Shandong Normal University, China}}
\affiliation{\institution{$^4$Shandong Sport University, China}}
\email{{xinxiao.zhao1985, jason.zy.cheng}@gmail.com, zhengjiecai@sdu.edu.cn}
\thanks{Corresponding author: Zhiyong Cheng and Jiecai Zheng}

\begin{abstract}
Recommender systems, which merely leverage user-item interactions for user preference prediction (such as the collaborative filtering-based ones), often face dramatic performance degradation when the interactions of users or items are insufficient. In recent years, various types of side information have been explored to alleviate this problem. Among them, knowledge graph (KG) has attracted extensive research interests as it can encode users/items and their associated attributes in the graph structure to preserve the relation information. In contrast, less attention has been paid to the item-item co-occurrence information (i.e., \textit{co-view}), which contains rich item-item similarity information. It provides information from a perspective different from the user/item-attribute graph and is also valuable for the CF recommendation models. In this work, we make an effort to study the potential of integrating both types of side information (i.e., KG and item-item co-occurrence data) for recommendation. To achieve the goal, we propose a unified graph-based recommendation model (UGRec), which integrates the traditional directed relations in KG and the undirected item-item co-occurrence relations simultaneously. In particular, for a directed relation, we transform the head and tail entities into the corresponding relation space to model their relation; and for an undirected co-occurrence relation, we project head and tail entities into a unique hyperplane in the entity space to minimize their distance. In addition, a head-tail relation-aware attentive mechanism is designed for fine-grained relation modeling. Extensive experiments have been conducted on several publicly accessible datasets to evaluate the proposed model. Results show that our model outperforms several previous state-of-the-art methods and demonstrate the effectiveness of our UGRec model.
\end{abstract}

\ccsdesc[500]{Information systems~Recommender systems}

\keywords{Graph-based, Metric Learning, Recommendation, Attention Mechanism}

\maketitle

\section{INTRODUCTION}
The recommender system plays an important role in tackling the information overload problem in many real systems, such as E-commerce websites and social platforms. Collaborative filtering has been the most widely studied recommendation technique, such as matrix factorization \cite{a2} and collaborative metric learning \cite{a11,a19}. Those models have made a great success due to their simplicity by only exploiting user-item interaction information for preference modeling. However, they often suffer serious performance degradation when the interactions are sparse (i.e., cold-start or data sparsity problem)~\cite{cheng2018ancf,a41,chenliang2019}.

A straightforward method to alleviate this issue is to exploit side information which provides additional knowledge about users and items~\cite{cheng2018aspect,dkn,lu2019online,liu2020a2}. Many approaches have been developed in this direction. Among them, knowledge-graph (KG) based methods have attracted increasing attentions in recent years, because users, items and their attributes can be mapped into the KG to model the complex relations among them, facilitating more accurate user preference modeling. Another advantage of the KG-based recommendation method is that the well-developed translating embedding models (such as TransE \cite{a15}, TransD \cite{a18}) for multi-relation modeling can be directly applied. In the real recommender systems, besides the explicit interactions between user and item that can be directly utilized for preference modeling, the rich item-item co-occurrence information is also useful for user preference modeling. The co-occurrence relation can be of users (e.g., \textit{friends}) or items (e.g., \textit{co-buy} and \textit{co-view}), which provides implicit user-user and item-item similarity information and is beneficial for user and item embedding learning. In fact, the item-item co-occurrence relations have been exploited to enhance the performance in several studies~\cite{a9,a25,a57}. More importantly, the user-item interactions are often sparse, such as the purchase behaviors in E-commerce websites, but the co-occurrence information is abundant (e.g., \emph{co-view}). Therefore, the co-occurrence information has a great potential to enhance the recommendation performance.

\begin{figure}[]
  \centering
  \includegraphics[width=\linewidth]{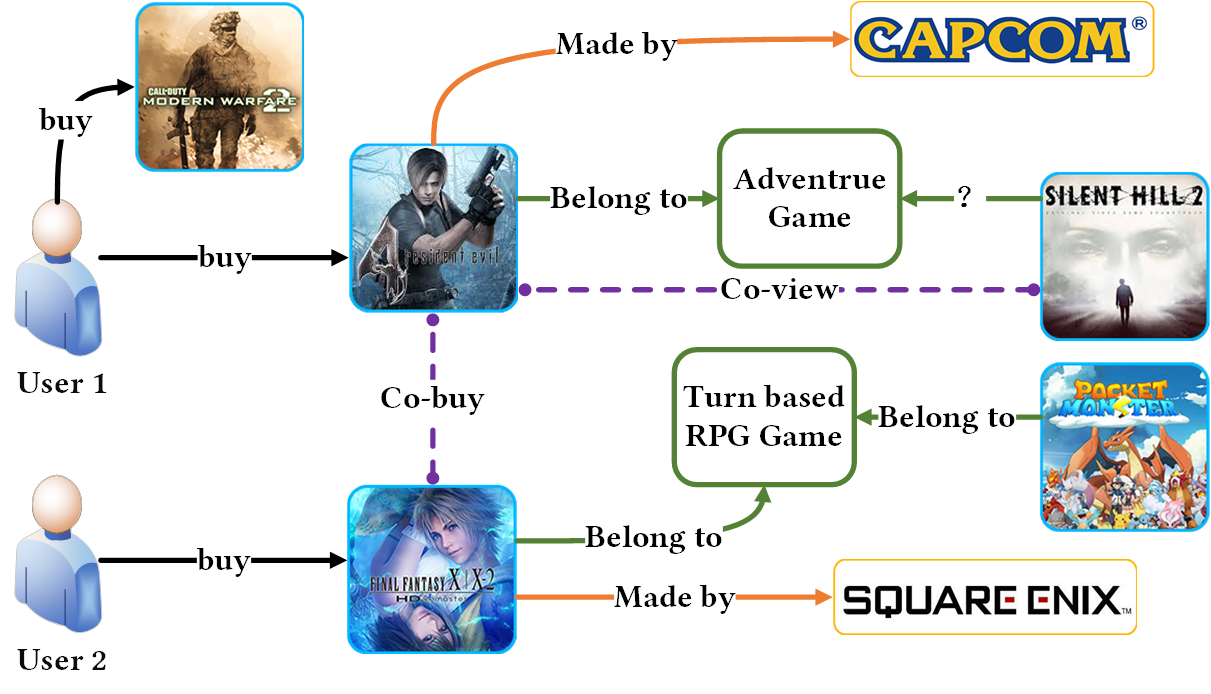}
  \caption{A toy example of our unified graph, which contains both directed and undirected relations. In the figure, the solid and dotted lines indicate the directed and undirected relation, respectively.}
  \label{fig:ex}
\end{figure}

Although both the aforementioned two types of information are valuable for recommendation and some methods have been developed to exploit them individually, they have been seldom used together in a unified recommendation model.  We claim that the co-occurrence information is complementary to the knowledge information in KG, and thus combining both types of information  can further improve the performance. The item-item co-occurrence relation contributes additional collaborative signals besides the user-item interactions, and the directed knowledge information (e.g., item attributes) often provides direct semantic features of items.  More concretely, we deem that the co-occurrence information can complement the KG from the following two aspects.
\begin{itemize}[leftmargin=*]
\item The co-occurrence encodes underlying relations that cannot be revealed by the directed relations in KG. For example, two items with a \textit{co-buy} relation because of their complementary functions (e.g., iPhone and Air Pods, hamburger and cola). This type of relation is hard to be captured by information in a KG but is quite common in the item-item co-occurrence relations. Besides, some items share certain common features and thus are favored by the same users. However, the latent relations cannot be revealed in a KG.
Take the graph in Figure \ref{fig:ex} as an example, \textit{Resident Evil 4} is an \textit{action/adventure} game manufactured by \textit{Capcom}, and \textit{Finally Fantasy 4} is a \textit{turn-based RPG} game released by \textit{Square Enix}. Two games are different from \emph{manufacturer}, \emph{game category}, and \emph{game style}. There is no clear relation between them from the attribute information, but the consumption trend shows that they are often purchased together by video game lovers. It implies that there are some common features shared by them to arouse the consumers’ interests to buy them together, but the shared features are hard to be observed in the directed relations of a KG.

\item The co-occurrence can bridge the missing relations caused by incomplete information in KG. For example, in Figure \ref{fig:ex}, both \textit{Resident Evil 4} and \textit{Silent Hill 2} belong to adventure/action games, however, \textit{Silent Hill 2} misses the category information in the dataset. As a result, we cannot construct the connection between them via the category information. Fortunately, they have a \textit{co-view} relation, which can be used to learn their relations and thus alleviates the category information missing problem.
\end{itemize}

The above observations motivate us to integrate the item-item co-occurrence information with the traditional KG for recommendation. However, the combination of both types of information is not trivial. Note that the relations in a KG are often directed and the co-occurrence relations are naturally undirected. A directed relation in a KG is commonly presented as subject-property-object triplet facts \cite{a16,a17}, i.e.,  $\langle h,\ t,\ r \rangle$, where each triplet describes a relation $r$ from head entity $h$ to tail entity $t$. The head and tail entities play different roles. For example, $\langle item\ i,\ category\ c,\ belong\ to\rangle$ represents the item $i$ belongs to the category $c$. In contrast, the head and tail entities in a co-occurrence relation are often of the same type and play the same role, e.g., $\langle item\ j,\ item\ k,\ co\text{-}buy \rangle$ is equivalent to $\langle item\ k,\ item\ j,\ co\text{-}buy \rangle$ under the \textit{co-buy} relation, in which both the head and tail are item entities. We noticed that CFKG~\cite{a20} exploits both the directed knowledge and undirected co-occurrence relations. However, this model simply applies the TransE\cite{a16} to model both the directed and undirected relations, which is not optimal and has the risk of generating a trivial solution in the optimization process, as discussed in~\cite{a9}. Given an co-occurrence triplet $\langle e_1,\ e_2,\ r \rangle$ and its reverse triplet $\langle e_2,\ e_1,\ r \rangle$, the use of TransE \cite{a15} will make $e_1+r_1 \approx e_2$ and $e_2+r_1 \approx e_1$ hold simultaneously, resulting in a trivial solution with $r_1 \approx 0$ and $e_1 \approx e_2$. It weakens the impact of relations on the head and tail entities.

To tackle the above limitation, in this work, we propose a unified graph-based recommendation (or \textbf{UGRec} for short) method, which adopts different strategies to learn the directed and undirected relations in an end-to-end manner. Specifically, we construct a heterogeneous graph, which contains both the directed knowledge relations and undirected item-item co-occurrence relations. Following the previous methods \cite{a20,a19,a11}, the user-item interaction behavior (e.g., purchase or click) is also considered as a kind of directed relation from the user to item. In our model, the directed and undirected relations are separately modeled, and all the entities are embedded into the same entity space. For each directed relation triplet, we apply TransD  \cite{a18} to map the head and tail entities from the entity space to the corresponding directed relation space, and then optimize the model based on a metric learning approach. For the undirected co-occurrence relation, we project the head and tail entities into a unique hyperplane in the entity space and then minimize their distance on this hyperplane. As discussed in \cite{a19,a38}, for those triplets of the same relation, a fixed relation embedding is insufficient to well model the relations of different head-tail entity pairs. To solve this problem, we design a head-tail relation-aware attention mechanism for fine-grained relation modeling. To evaluate the effectiveness of our UGRec model on recommendation, we conduct extensive experiments on three real-world datasets. The results show that our model outperforms the several strong baselines by a large margin, demonstrating the potential of combining both knowledge and co-occurrence information in recommendation. In addition, we also  perform an ablation study to examine the effectiveness of different components in our UGRec model.

In summary, the main contributions of this work are as follows:
\begin{itemize}[leftmargin=*]
\item We discuss the complementary nature of the undirected item-item co-occurrence and the directed knowledge information for modeling user preference, and emphasize the potential of exploiting both types of information  for recommendation.
\item We propose a unified graph-based recommendation method called UGRec\footnote{Codes are available at https://github.com/garfieldcat1985/ugrec.}, which can effectively leverage the undirected co-\linebreak occurrence and directed knowledge information in a graph for user preference modeling. In particular, UGRec models the directed relations and undirected relations with different strategies; and a head-tail relation-aware attention mechanism is designed to model the fine-grained relation for each triplet in the graph.
\item We conduct extensive experiments on three public benchmarks, demonstrating the superiority of our UGRec model on exploiting both the directed and undirected relations for recommendation.
\end{itemize}

\section{Our Proposed Model}
\subsection{Preliminaries}
\subsubsection{Problem Setting} Given a dataset consists of a user set $\mathcal{U}$, an item set $\mathcal{I}$, a knowledge or attribute (e.g., \emph{manufacturer}, \emph{category}) set $\mathcal{K}$, a user interaction matrix $\mathcal{Y}^{\mathcal{N}_u\times\mathcal{N}_i}$, as well as an item-item co-occurrence relation set $\mathcal{C}$. $\mathcal{N}_u$ and $\mathcal{N}_i$ are the numbers of users and items, respectively. For each $y_{(u,i)}\in \mathcal{Y}$, $y_{(u,i)}=1$ indicates the user $u$ has interacted with the item $i$ (such as \emph{click}); otherwise, $y_{(u,i)}=0$. In addition, a directed relation set $\mathcal{R}_d$ can be constructed based on the attribute set $\mathcal{K}$. Each directed relation is represented by a triplet $\langle h,\ t,\ r_d  \rangle$, where $h$ and $t$ denote the head entity and tail entity, respectively; $r_d$ denotes the directed relation from $h$ to $t$. For example, $\langle Avatar,\ James Cameron,\
directed\ by \rangle$ means that the movie  \textit{Avatar} is directed by \textit{James Cameron}. Note that each interaction between a user and an item is also regarded as a directed relation in this work. In particular, the interaction behavior between a user $u$ and an item $i$ is represented as $\langle u,\ i,\ r_{inter} \rangle$ or $\langle h,\ t,\ r_{inter} \rangle$ in our context with $r_{inter} \in \mathcal{R}_d$ as a type of directed relation. Similarly, an undirected relation set $\mathcal{R}_c$ is constructed based on the co-occurrence relation set $\mathcal{C}$. Each undirected relation is represented by a triplet $\langle h,\ t,\ r_c \rangle$; where $r_c$ is an undirected relation between two items $h$ and $t$. For an undirected relation,  $\langle h,\ t,\ r_c \rangle$ is equivalent to $\langle t,\ h,\ r_c \rangle$. Based on the above definitions, we define a unified graph $\mathcal{G}=(\mathcal{E},\mathcal{R})$,  where $\mathcal{E}=\mathcal{U}\cup \mathcal{I}\cup \mathcal{K}$  and $\mathcal{R}=\mathcal{R}_d\cup \mathcal{R}_c$ are the node set and edge set of the graph, respectively. To this end, the graph contains all the users, items, attributes as well as the directed and undirected relations. The directed and undirected relations are represented as directed and undirected edges in the graph.

In this paper, we target at the top-$n$ recommendation, namely, recommending a target user a list of items, which have not been consumed by the user yet and most likely to be appealing to the user. As  the interaction relations between users and items are denoted as directed relations/edges in our model, \emph{the preference prediction of a user towards an item} is converted to \emph{a relation prediction from a user $u$ to item $i$}, which is also denoted as \emph{the relation triplet   $\langle h,\ t,\ r_{inter} \rangle$}. Our model aims to capture both the directed and undirected relations among users, items, and attributes for the relation prediction between users and items. In the following presentation, we will use head entity $h$, tail entity $t$, and relation $r$ to denote all the entities and relations in the graphs. Notice that the users and items are also contained in the head and tail entities, and we will not distinguish the different entities and relations in the following description.

\begin{figure*}[htbp]
\centering
\includegraphics[width=0.9\linewidth]{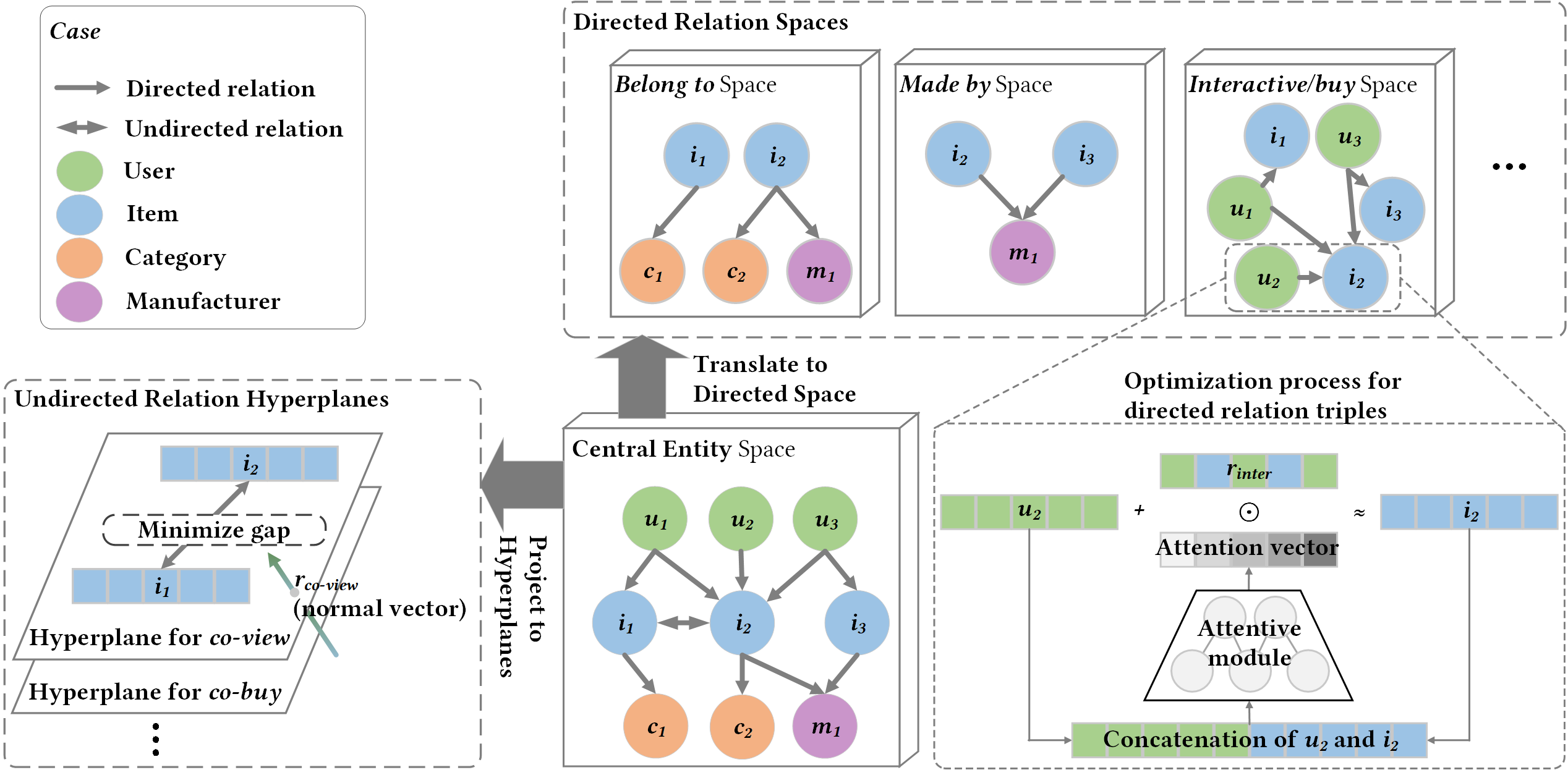}

\caption{An illustration of our UGRec model: 1) The directed relations are translated into the corresponding relation space for optimization; 2) the undirected co-occurrence relations are optimized on the hyperplanes projected from the entity space.}
\label{fig:2}
\end{figure*}

\subsection{Our UGRec Model}
In our proposed UGRec model, the two types of information are represented as directed and undirected relations respectively in a unified graph. As discussed, the traditional translating embedding methods are not suitable for modeling the undirected relations. Therefore, for a better performance, we use two different methods to model the directed and undirected relations separately in the UGRec model. In the next, we elaborate the directed relation modeling and undirected modeling sequentially. As they are described separately in two subsections, for ease of presentation, we will use $r$ to denote the relation without specifying $r_d$ or $r_c$ in the following description.

\subsubsection{Directed Relation Modeling}
For the directed relation modeling, we follow the traditional translating embedding models, such as SE \cite{a29} and TransD \cite{a18}, to construct a hierarchical vector space, which contains a central entity space  and multiple directed relation spaces. Each type of directed relation corresponds to a specific relation space.  In our work, the dimension of all vector spaces is set to $k$. Each entity (head or tail) is represented as an embedding vector in the central entity space, i.e.,  $\mathbf{h}\in \mathbb{R}^{k \times 1}$ or $\mathbf{t} \in \mathbb{R}^{k \times 1}$.\footnote{In the paper, unless otherwise specified, notations in bold style denote matrices or vectors, and the ones in normal style denote scalars.} Similarly, each directed  relation $r \in \mathcal{R}_d$ is also represented as an embedding vector $\mathbf{r}\in \mathbb{R}^{k \times 1}$.

\textbf{Relation modeling.} Following the common setting in translating embedding methods~\cite{a15,a18},  for a directed relation $\langle h,\ t,\ r \rangle$, we model the relation $r$ between the head entity $h$ and tail entity $t$ by optimizing $\mathbf{h}_{r}^d+\mathbf{r} \approx \mathbf{t}_{r}^d$. This is achieved by minimizing the distance between $\mathbf{h}_{r}^d$ and $\mathbf{t}_{r}^d$, which is defined as:

\begin{equation}\label{eq:direct_minimize}
{f}_{r}^d(h,t)=\left\| \mathbf{h}_{r}^d+\mathbf{r}-\mathbf{t}_{r}^d \right\|_2^2
\end{equation}
where $\mathbf{h}_{r}^d$ and $\mathbf{t}_{r}^d$ are the projection embeddings of the head entity and tail entity in the space of the relation $r$. In other words, the modeling of a directed relation $\langle h,\ t,\ r \rangle$ is to minimize the distance between the projections of head entity $h$ and tail entity $t$ in the corresponding relation space. The projection embeddings $\mathbf{h}_{r}^d$ and $\mathbf{t}_{r}^d$ are obtained by:
\begin{equation}
\left\{
\begin{aligned}
\mathbf{h}_{r}^d&=\mathbf{M}_{rh} \mathbf{h}\\
\mathbf{t}_{r}^d&=\mathbf{M}_{rt} \mathbf{t}
\end{aligned}
\right.
\end{equation}
where $\mathbf{M}_{rh}$ and $\mathbf{M}_{rt}$ are the mapping matrices for $\mathbf{h}_{r}^d$ and $\mathbf{t}_{r}^d$, respectively. Rather than simply using a randomly-initialized matrix as the translator between an entity space and the target directed relation space, we apply the strategy used in TransD~\cite{a18}, which has been demonstrated to be capable of simplifying the translation operation and alleviate potential time-consuming and overfitting issues. Specifically, the mapping matrices are obtained by:
\begin{equation}
\left\{
\begin{aligned}
\mathbf{M}_{rh}&=\mathbf{r_p} \mathbf{h_p^{\top}}+\mathbf{I}^{k\times k}\\
\mathbf{M}_{rt}&=\mathbf{r_p} \mathbf{t_p^{\top}}+\mathbf{I}^{k\times k}
\end{aligned}
\right.
\end{equation}
where $\mathbf{h_p}$, $\mathbf{t_p}$, and $\mathbf{r_p}$ are unique projection vectors for the head entity $h$, tail entity $t$, and directed relation $r$, respectively. $\mathbf{I}^{k\times k}$ denotes an identity matrix, which can bring us a better performance in practice. They are also the embedding vectors to be learned in the model. ${\top}$ is the transpose operator. From the equation, we can see that for the mapping of an head entity $h$ to a relation space $r$, the unique projection vectors (e.g., $\mathbf{h_p}$ and $\mathbf{r_p}$) of both $h$ and $r$ are all involved into the mapping matrix computation. In this way, for each entity, a customized entity-relation aware mapping matrix is employed to project it into the target directed relation space. This can represent an entity differently on different directed relation spaces~\cite{liu2019user}.

\textbf{Head-tail relation-aware attention.} The above method models the multiple directed relations on a global level. However, we argue that it cannot distinguish the subtle difference inside a relation between different \emph{head-tail}  entity pairs. Taking the user-item \emph{interaction} relation as an example, two users (denoted as $h$ and $\hat{h}$) may interact with the same item $t$ because of different reasons.  For instance, a user likes a video game because of its beautiful scenery while another user is attracted by its interesting storyline. In this case, although both users connect to the item based on the same relation, they relate to it because of different reasons. For the two triplets $\langle h,\ t,\ r \rangle$ and $\langle \hat{h},\ t,\ r \rangle$ (here we use $r$ instead of $r_{inter}$ for simplicity), both triplets are modeled as $\mathbf{h}_{r}^d+\mathbf{r} \approx \mathbf{t}_{r}^d$ and $\mathbf{\hat{h}}_{r}^d+\mathbf{r} \approx \mathbf{t}_{r}^d$, which cannot differentiate the reason that why a user relates to the item (i.e., the subtle difference on the relation). Motivated by the above consideration, we design a head-tail relation-aware attention mechanism to model the fine-grained relation for each triplet, making our model capable of distinguishing the differences between the relations of different triples inside a relation embedding. Given a directed relation triplet $\langle h,\ t,\ r \rangle$, the attention vector $\mathbf{Att}^d_{(\mathbf{h}_{r}^d, \mathbf{t}_{r}^d)}$ for the head to tail relation is computed  as:
\begin{equation} \label{eq:att1}
\mathbf{z}=ReLU(\mathbf{W}[\mathbf{h}_{r}^d:\mathbf{t}_{r}^d]+\mathbf{b})
\end{equation}
\begin{equation} \label{eq:att2}
Att^d_{(\mathbf{h}_{r}^d, \mathbf{t}_{r}^d,l)}=\frac{exp(\mathbf{z}_l)}{\sum_{i=0}^k exp(\mathbf{z}_i)}
\end{equation}
where $\mathbf{W}$ and $\mathbf{b}$ are trainable weight matrix and bias vector,  respectively. $\mathbf{h}_{r}^d$ and $\mathbf{t}_{r}^d$ are the mapping vectors of $h$ and $t$ in the $r$ relation space. $[\mathbf{h}_{r}^d:\mathbf{t}_{r}^d]$ denotes a concatenation of the two vectors $\mathbf{h}_{r}^d$ and $\mathbf{t}_{r}^d$. \emph{ReLU}~\cite{maas2013rectifier} is used as the activation function.
By integrating the obtained attention vector into Eq.\ref{eq:direct_minimize}, the distance function becomes:
\begin{equation} \label{eq:ddis}
{f}_{r}^d(h,t)=\left\| \mathbf{h}_{r}^d+\mathbf{r} \odot \mathbf{Att}^d_{(\mathbf{h}_r^d,\mathbf{t}_r^d)} -\mathbf{t}_{r}^d \right\|_2^2
\end{equation}

\textbf{Optimization.} Following the common optimization strategy of the metric-learning based approaches \cite{a11,a19}, we adopt the pairwise learning strategy for optimization. For each directed relation $r$, given the positive triplet set $\mathcal{S}_{r}$ and the negative triplet set $\overline{\mathcal{S}_{r}}$, the loss function to guide the optimization process is defined as:
\begin{equation} \label{eq:dop}
\mathcal{L}_{r}^D =\sum_{\langle h,\ t,\ r \rangle \in \mathcal{S}_{r}} \sum_{\langle h,\ t',\ r \rangle \in \overline{\mathcal{S}_{r}}} [m+ \mathit{f}_{r}^d(h,t) -\mathit{f}_{r}^d(h,t') ]_+
\end{equation}
where $[\cdot]_+$ denotes the standard hinge loss, and $m>0$ is the safety margin size. Notice that for each relation $ r \in \mathcal{R}_d$, we can compute a loss $\mathcal{L}_{r}^D$. The loss of all types of relation are aggregated to obtain the final loss (see Section~\ref{sec:learn} for details).

\subsubsection{Undirected Relation Modeling}
An undirected co-occurrence relation between two entities often implies that they share some common features (e.g., the \emph{co-view} relations between two items). The key of modeling an undirected relations for a triplet  $\langle h,\ t,\ r \rangle$ is to keep the equivalence between $f_{r}(h,t)$ and $f_{r}(t,h)$ and avoid the trivial solution in the optimization process (see the discussion in Section 1). To meet those needs, a straightforward method is to adopt the DistMult strategy as in~\cite{a9,a21} to model the similarity between two entities with an undirected relation:
\begin{equation} \label{eq:cdis}
f_{r}^c(h,t)= f_{r}^c(t,h)=\mathbf{t}^{\top} \cdot diag(\mathbf{r} ) \cdot \mathbf{h}
\end{equation}
where $diag(\mathbf{r})$ denotes a diagonal matrix with the diagonal elements equal to $\mathbf{r}$. The stand BPR loss function~\cite{a5} can be used to minimize $f_{r}^c(h,t)$ with the positive and negative sampling strategy, such as in~\cite{a9}. However, this method does not perform well in our experiments. The main reason is that it uses the  dot product to measure the similarity between head and tail entities (see Eq.~\ref{eq:cdis}), which is incompatible with the metric-learning approach used in the translating embedding models.

\textbf{Relation modeling.} Based on the above observation, we propose a metric-learning based method to substitute the DistMult strategy to model the undirected relations in our model.  Notice that although a co-occurrence relation implies some common features are shared by the head and tail entities, they are still different entities. Besides, directly minimizing their distance in the entity space may also suffer from the risk of a trivial solution. With those considerations, for an undirected triplet $\langle h,\ t,\ r \rangle$, we first project the two entities $h$ and $r$ to a hyperplane for the relation $r$; and then minimize the distance between the two entities on this hyperplane instead of in the original entity space. The hyperplane is unique for each triplet $\langle h,\ t,\ r \rangle$ to minimize the distance between the head and tail entities. In addition, lessening the optimization constraint from the entity space to a hyperplane is beneficial to make entity embeddings more uniformly distributed in the entity space. Under this setting, the projection vectors $\mathbf{h}_{r}$ and $\mathbf{t}_{r}$ of $h$ and $t$ to $r$'s hyperplane are obtained by:
\begin{equation} \label{eq:hyperplane}
\left\{
\begin{aligned}
\mathbf{h}_{r}^c&=\mathbf{h}-\frac{{\mathbf{r}}^\top  \cdot \mathbf{h} \cdot \mathbf{r}}{\left\| \mathbf{r}\right\|_2^2}    \\
\mathbf{t}_{r}^c&=\mathbf{t}-\frac{{\mathbf{r}}^\top \cdot \mathbf{t}  \cdot \mathbf{r}}{\left\| \mathbf{r} \right\|_2^2}
\end{aligned}
\right.
\end{equation}
The distance function $f_{r}^c$ for  $r$ relation triplets is defined as:
\begin{equation}
f_{r}^c(h,t)=f_{r}^c(t,h)=\left\| \mathbf{h}_{r}^c-\mathbf{t}_{r}^c   \right\|_2^2
\end{equation}

 Because the hyperplane is relation-aware and unique for each head-tail pair, we directly minimize the distance between $\mathbf{h}_{r}^c$ and $\mathbf{t}_{r}^c$ after projecting the head entity $h$ and tail entity $t$ to the hyperplane.  This is different from that in TransH~\cite{a16}, which employs another vector as the relation vector to minimize their distance.

\textbf{Head-tail relation-aware attention.} Similar to directed relation modeling, we also adopt an attentive mechanism for fine-grained relation modeling. For simplicity, the attentive module  adopted here is the same as the one used in the directed relation. The attention vector $\mathbf{Att}^c_{(\mathbf{h}_r^c, \mathbf{t}_r^c)}$for a undirected triplet $\langle h,\ t,\ r \rangle$  is computed exactly the same as the one described in Eq.~\ref{eq:att1} and Eq.~\ref{eq:att2}. With the attention vector, Eq.~\ref{eq:hyperplane} becomes:

 \begin{equation}
\left\{
\begin{aligned}
\mathbf{\hat{r}}&=\mathbf{r} \odot \mathbf{Att}^c_{(\mathbf{h}_r^c, \mathbf{t}_r^c)} \\
\mathbf{h}_{r}^c&=\mathbf{h}-\frac{{\mathbf{\hat{r}}}^\top  \cdot \mathbf{h} \cdot \mathbf{\hat{r}}}{\left\| \mathbf{\hat{r}}\right\|_2^2}    \\
\mathbf{t}_{r}^c&=\mathbf{t}-\frac{{\mathbf{\hat{r}}}^\top \cdot \mathbf{t}  \cdot \mathbf{\hat{r}}}{\left\| \mathbf{\hat{r}} \right\|_2^2}
\end{aligned}
\right.
\end{equation}

\textbf{Optimization.} We adopt the same method as the one used for the directed relations (see Eq.~\ref{eq:dop}) for optimization.  To save space, we omit the details here for computing the loss $\mathcal{L}_{r}^C$ for each undirected relation $r\in \mathcal{R}_c$.

\subsection{Model Learning and Recommendation} \label{sec:learn}
To effectively learn the parameters for our UGRec model, we aggregate the loss of all the directed and undirected relations in a unified manner through a multi-task learning framework. The total object function of UGRec is defined as:
\begin{equation}
\begin{aligned}
& \min_{\Theta} ~ \mathcal{L} =\lambda_d \sum_{r_i \in \mathcal{R}_d}\mathcal{L}_{r_i}^D +\lambda_c \sum_{r_j \in \mathcal{R}_c} \mathcal{L}_{r_j}^C \\
& \textrm{s.t.} ~ ||\mathbf{h}||_2 \leq 1, ||\mathbf{t}||_2  \leq 1, ||\mathbf{r}||_2  \leq 1, ||\mathbf{h_p}||_2  \leq 1, ||\mathbf{t_p}||_2  \leq 1, ||\mathbf{r_p}||_2  \leq 1 \\
\end{aligned}
\end{equation}
where $\Theta$ is the total parameter space, including all embeddings and the variables of attention networks. Different from the traditional $L_2$ regularization, we constrain all the embeddings within a Euclidean unit sphere, which has been demonstrated to be effective in both knowledge embedding~\cite{a9,a18} and recommendation~\cite{a16,a33,liu2019user}. This constraint can also effectively prevent the risk of a trivial solution, which is simply scaling up the norm of the corresponding embeddings to minimize the loss (both $\mathcal{L}_{r}^D$ and $\mathcal{L}_{r}^c$). $\lambda_d$ and $\lambda_c$ are hyperparameters to control the contributions of the undirected and directed relations. Notice that we can also further fine-tune the contribution of some important relations (e.g., the interaction relation). As our focus is to study the effects of combining both types of relations for recommendation, we have not put much efforts on tuning those hyperparameters and simply set them equally (i.e., $\lambda_d=\lambda_c=1$). We leave the study of the impact of different relations on the recommendation accuracy as a further work.   The AdaGrad algorithm~\cite{duchi2011adaptive} is used to optimize the whole model.

\textbf{Recommendation.} After training the model, we can obtain the embeddings of all the node entities in the graph, which include the user and item entities. For a target user $u$, the recommendation is achieved by estimating the similarity scores of all the items that she has not consumed before (denoted by  $\mathcal{I}_{\overline{u}}$) in the space of the directed relation  \emph{``interaction"}. More specifically, for each item $i$ in $\mathcal{I}_{\overline{u}}$, we compute its distance to the user  $f_{r_{inter}}^d(u,i)$ based on the Eq.~\ref{eq:ddis}. The top-$n$ items with the shortest distances are regarded as the most similar ones and recommended to the user.

\section{EXPERIMENTS}
\begin{table*}
  \caption{Basic statistics of the experimental datasets, \#user, \#item and \#interactions represent the number of users and items and interactions; \#r$_x$ indicates the number of corresponding relation triplets.}
  \label{tab:dataset}
  \setlength{\tabcolsep}{2.8mm}
  \begin{tabular}{cccccccccc}
    \toprule
    Dataset & \#user & \#item & \#interact & sparsity & \#r$_{belong\_to}$ & \#r$_{made\_by}$ & \#r$_{co\_buy}$ & \#r$_{co\_view}$ & \#r$_{total}$\\
    \midrule
    Games &	91,568 & 21,820 & 620,361 &	99.96$\%$ &	40,654 & 18,142 & 133,291 & 82,708 & 274,795\\
    CDs \& Vinyls & 41,690 & 32,601 &	338,296 & 99.98\% &	65,271 & 29,330 & 62,962 & 31,797 & 189,360\\
    Movies \& TVs & 33,680 & 25,185 & 257,432 & 99.97\% & 54,018 & 21,731 & 48,130	 & 24,225 &	148,104\\
    \bottomrule
  \end{tabular}
\end{table*}

To validate the effectiveness of our model, we conducted extensive experiments on three public datasets to answer the following research questions:
\begin{itemize}[leftmargin=*]
\item RQ1: Does our proposed UGRec outperforms the state-of-the-art (SOTA) methods on the top-$n$ recommendation task?
\item RQ2: Can our model benefit from the combination of the co-occurrence and knowledge relations and achieve better performance  than the SOTA KG-based models?
\item RQ3: Can our model effectively leverage the co-occurrence and knowledge relations to alleviate the data sparsity problem?
\item RQ4: How do different components (such as the directed and undirected relations, the attention mechanism) of our model impact the final performance?
\end{itemize}

\subsection{Experiment Setup}
\subsubsection{Datasets}
We adopt the widely-used Amazon dataset\footnote{http://deepyeti.ucsd.edu/jianmo/amazon/index.html.} to evaluate our model in experiments. In our experiments, we adopt three product categories-\textit{Games}, \textit{CDs \& Vinyls}, \textit{Movies \& TVs}, and retain users and items with at least 4 interactions. For each dataset, we extract knowledge relation triplets to evaluate the knowledge-aware recommendation models. Specifically, we extracted two directed relations: \textit{belong to} and \textit{made by}; for the co-occurrence relation, the \textit{co-buy} and \textit{co-view} information and adopted.\footnote{In this paper, we just study the potential of item side co-occurrence relation modeling, and leave the study for user side co-occurrence relation modeling to future works.} The statistic information of the adopted datasets is summarized in Table~\ref{tab:dataset}.

\subsubsection{Evaluation protocols.} The standard \textit{leave-one-out} strategy~\cite{a5} for recommendation evaluation is adopted. Specifically, the latest interaction of each user is held-out as the testing data, the second latest interaction is reserved as the validation data, and the remaining interactions are used for training.  For each user, all the items which are not interacted by this user are used as testing items with the held-out positive item. In the testing stage, all the items are ranked based on the adopted recommendation model to evaluate the performance. Two standard evaluation metrics - hit ratio value (HR) and normalized discounted cumulative gain value (NDCG) are employed in evaluation.
\subsubsection{Baselines} To demonstrate the effectiveness, we compare our proposed UGRec with several competitive recommendation models based on different approaches.
\begin{itemize}[leftmargin=*]
\item NeuMF \cite{a3}: This model generalizes the matrix factorization to neural networks. It takes advantage of the neural layers to capture the non-linear interactions between users and items.
\item LightGCN \cite{a44}: This model uses only the user-item interaction data, and inherits the metric of GCN-based methods by exploiting the high-order proximities and simplifies the recent NGCF model~\cite{a40} by removing the nonlinear activation and feature transformer module.

\item CML \cite{a11}: This method also only exploits the user-item interaction data. It replaces the preference prediction function (i.e., dot product) with a metric learning approach (i.e., Euclidean distance). The items with shortest distance to the target user in the embedding space are recommended to the user.
\item LRML \cite{a19}: This method introduces a memory-based attention module into CML to model latent relations between users and items. This module can effectively tackle the geometric inflexibility issue of CML.
\item CFKG \cite{a20}: This model treats the user-item interaction as a directed relation as other knowledge relation triplets, and employs the TransE methods to learn the entity embeddings.
\item KTUP \cite{a28}: This is a recently proposed translation-based recommendation model based on KG. In particular, it unifies the recommendation task with a KG completion model to enjoy the mutual enhancements, achieving state-of-the-art performance.
\end{itemize}
Among these methods, NeuMF, LightGCN, CML and LRML exploit only the user-item interaction data. The last two methods all employ KG embedding learning techniques to learn the latent relations between entities, notice that they exploit the undirected co-occurrence relations in the same manner as the directed relations.
\begin{table*}
  \caption{Overall performance comparison on three datasets. Noticed that the values are reported by percentage with ’\%’ omitted.}
   \label{tab:result}
  \setlength{\tabcolsep}{4mm}
  \vspace{-5pt}
  \begin{threeparttable}
  \begin{tabular}{c|c||cc||cc||cc||c}
    \hline
    \multicolumn{1}{c|}{Datasets}&\multicolumn{1}{c||}{Metrics}&\multicolumn{1}{c}{NeuFM}&\multicolumn{1}{c||}{LightGCN}&\multicolumn{1}{c}{CML}&\multicolumn{1}{c||}{LRML}&\multicolumn{1}{c}{CFKG}&\multicolumn{1}{c||}{KTUP}&\multicolumn{1}{c}{UGRec}\\
    \hline\hline
    \multirow{2}{*}{Games}&HR@20&20.22&20.64&22.91&23.00&\textbf{24.03}&23.47&\textbf{24.92*}\\
    &NDCG@20&9.60&9.38&13.40&13.38&\textbf{13.51}&13.16&\textbf{13.99*}\\
    \hline
    \multirow{2}{*}{CDs \& Vinyls}&HR@20&17.98&18.13&19.83&19.89&21.77&\textbf{21.84}&\textbf{23.01*}\\
    &NDCG@20&8.56&9.11&12.13&12.19&11.93&\textbf{12.24}&\textbf{12.57*}\\
    \hline
    \multirow{2}{*}{Movies \& TVs}&HR@20&20.18&20.90&21.16&21.21&23.52&\textbf{23.85}&\textbf{24.79*}\\
    &NDCG@20&9.73&10.99&14.40&14.41&14.34&\textbf{14.43}&\textbf{14.60*}\\
    \hline
  \end{tabular}
  \begin{tablenotes}
    \footnotesize
        \item The symbol * denotes that the improvement is significant with p-value<0.05 based on a two-tailed paired t-test.
  \end{tablenotes}
  \end{threeparttable}
\end{table*}

\begin{figure*}
\subfigure[Games]{\includegraphics[width=.3\linewidth]{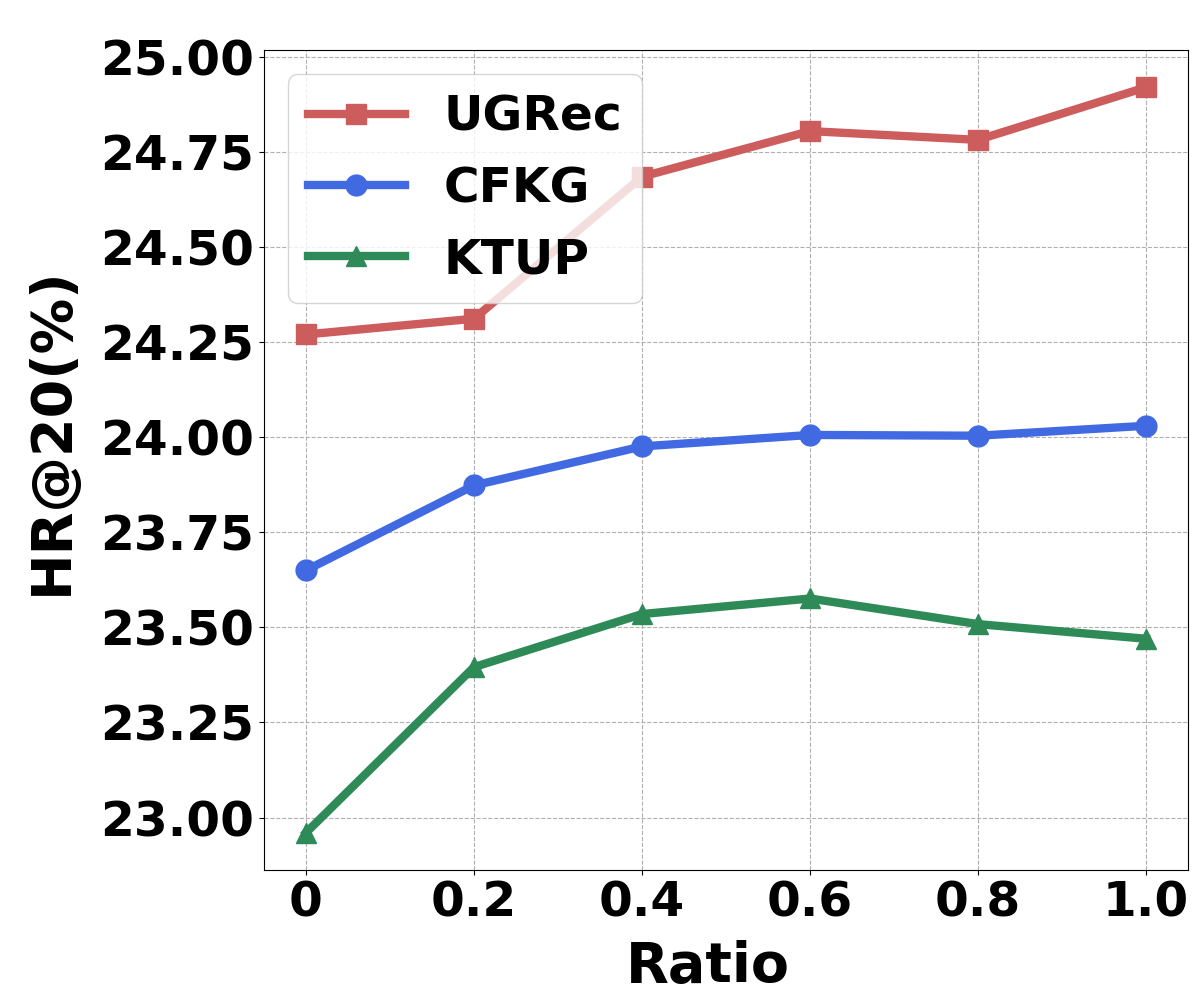}}
\subfigure[CDs \& Vinyls]{\includegraphics[width=.3\linewidth]{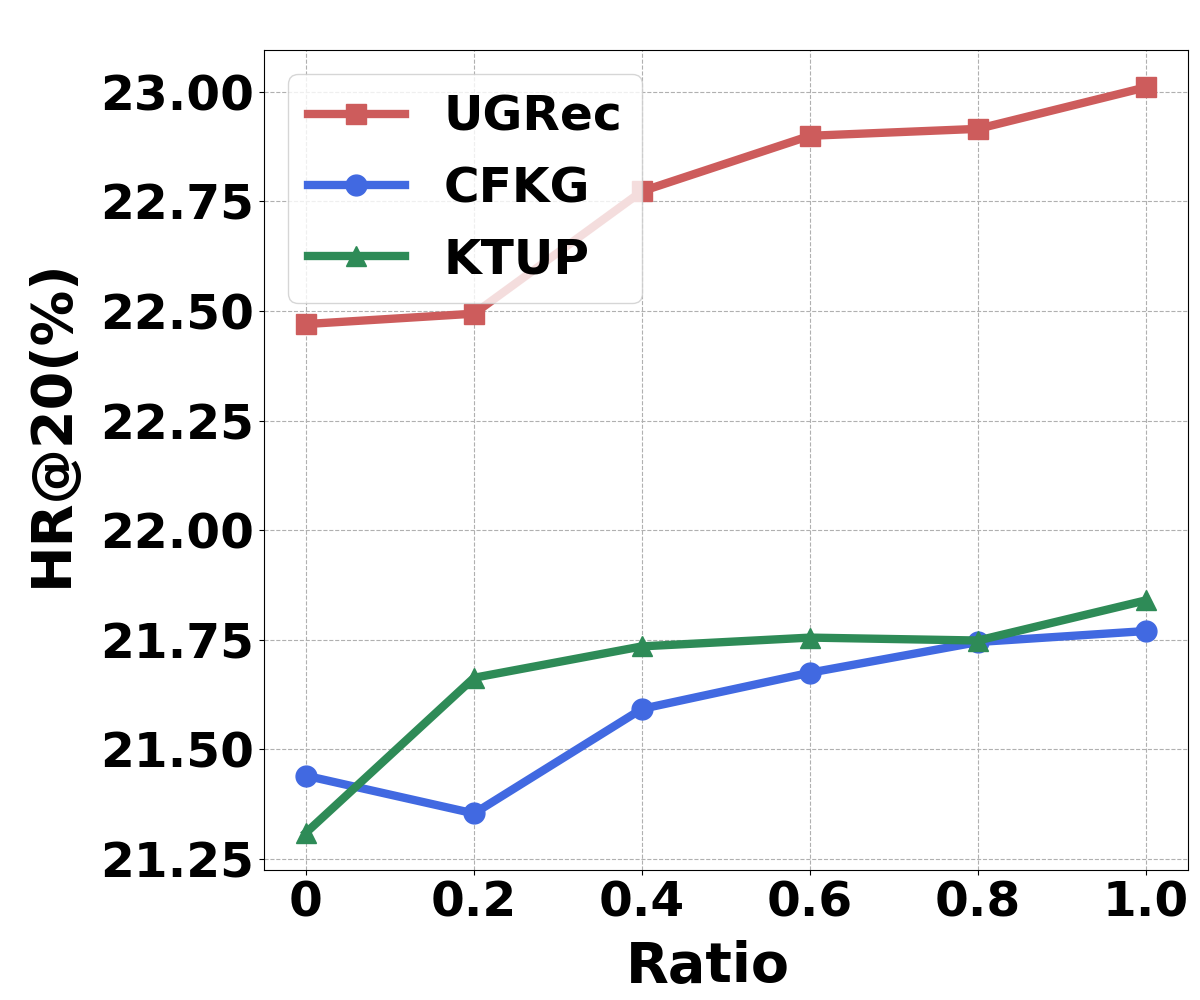}}
\subfigure[Movies \& TVs]{\includegraphics[width=.3\linewidth]{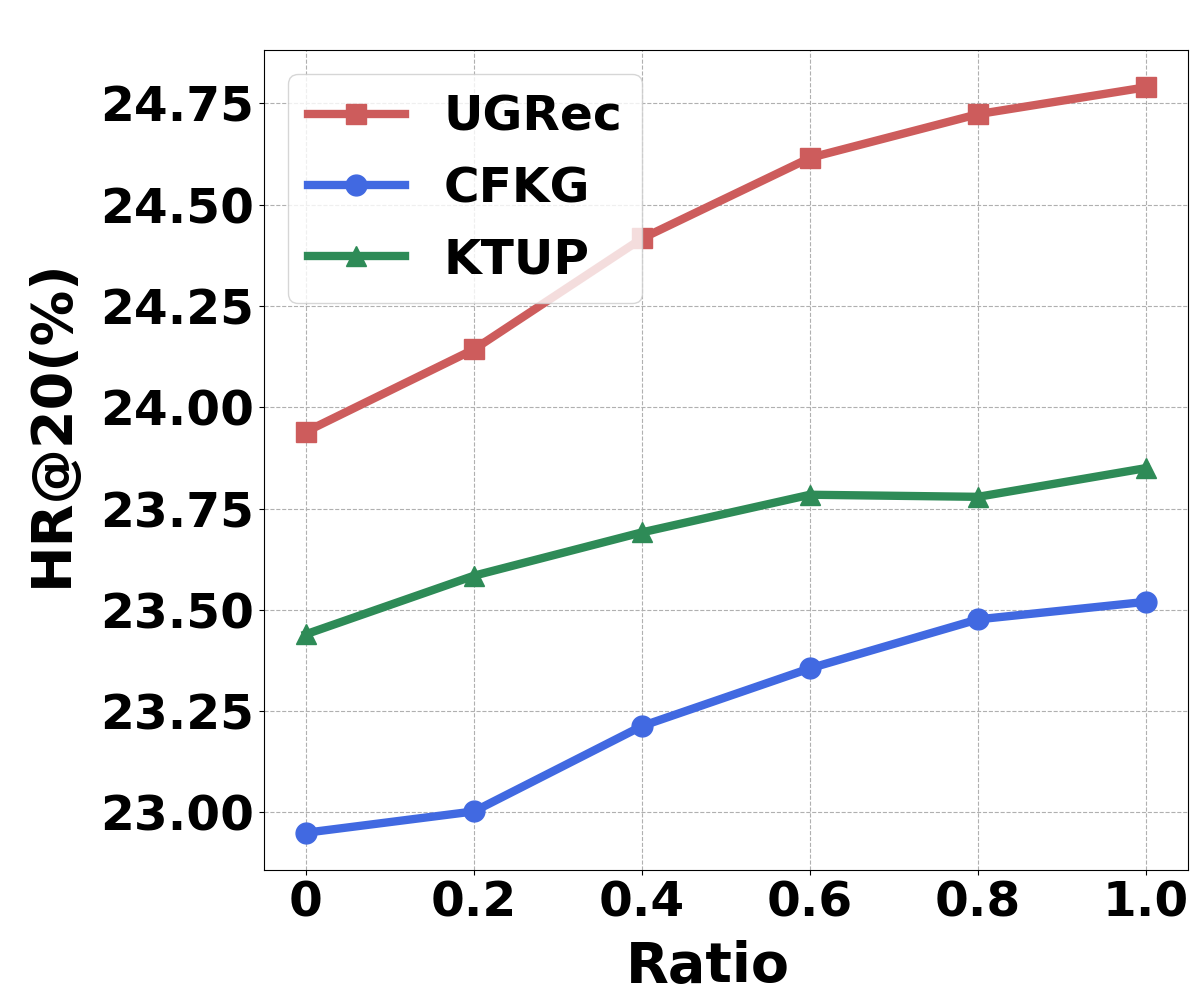}}
\vspace{-15pt}
\caption{Performance (HR@20) \emph{w.r.t.} different ratios of adopted co-occurrence triplets in the training data.}
\label{fig:co}
\end{figure*}

\subsubsection{Parameter Settings} \label{sec:parametersetting}
We implement our model with TensorFlow\footnote{https://www.tensorflow.org.} and put great efforts to tune hyper-parameters for all competitors. Specifically, for those models using hinge loss and BPR loss, we adopt the negative sampling strategy in \cite{a11,a18}, which first randomly samples $N$ negative instances and then select the hardest negative one to pair with the targeted positive instance. The hardest negative instance is the one which has the largest distance from the positive instance. In addition, to save the sampling time, the number of negative samples is bounded to 20 (i.e., $N=20$). For all models, we search the initial learning rate among $\{0,001,0.005,0.01,0.05,0.1\}$. The weights for the $L_2$ normalization in all models are searched among $\{{10}^{-4},{10}^{-3},{10}^{-2},{10}^{-1},\linebreak[0]{10}^0,{10}^1\}$. The margin value $m$ for the models using hinge loss are searched in the range of $\{1.1,1.2,...,2.0\}$ on these three datasets. In our experiments, $m=1.8$ (1.5 for \textit{Games}) works well. For our UGRec model, setting $m=1.8$ for the interaction relation, and $m=1$ for all other directed and undirected relations brings a better performance. For LRML, the optimal number of memory block number is 20; and for KTUP, the parameter $\lambda$ to balance TUP and TransH modules is set to 0.7 for the best performance. The propagation layer number of LightGCN is set to 2.  Finally, the dimension of the embedding vectors for users and items in all models is set to 64. The number of training epochs for all models is set to 1000, and their performance will be evaluated every 10 epochs.
\subsection{Experimental Results}
\subsubsection{Performance comparison (RQ1)} Table \ref{tab:result} reports the results of all the considered methods. Note that the best and second-best results are highlighted in bold. From the results, we have some valuable observations. Firstly, both NeuMF and LightGCN employ dot product as the similarity function for preference prediction, the better performance of LightGCN over NeuMF demonstrates the potential of exploiting high-order relations between users and items.  Secondly, CML and LRML achieve much better performance than NeuMF and LightGCN, which  validates the advantage of the metric learning based methods over the dot product based ones on modeling user preference in recommendation. Besides, LRML extends CML by introducing a relational vector to connect the user and item in the embedding space, achieving a better performance. This strategy has also been adopted in KTUP.

In the next, we discuss the performance of the methods with side information. CFKG, KTUP and our UGRec method all employ additional information besides the user-item interactions in the modeling. In general, the two KG-based methods (i.e., CFKG, KTUP) consistently outperforms the ones only exploiting user-item interaction data, demonstrating the benefits of incorporating additional information in recommendation. Our UGRec can further improve the performance by using different strategies to modeling the undirected and directed relations. UGRec outperforms all the baseline models consistently over all the datasets. Compared with the strongest baseline in terms of HR@20, UGRec can achieve a relative improvement by 3.70\%, 5.36\%, 3.94\% on \textit{Games}, \textit{CDs \& Vinyls} and \textit{Movies \& TVs}, respectively. Note that CFKG and KTUP also exploit both the directed knowledge and undirected co-occurrence data in our experiments, however, they model the undirected relations in the same way as the directed ones. As discussed, the co-occurrence is not suitable to be modeled as directed relations. In our UGRec model, we treat the directed  and undirected relations differently and obtain better performance.\footnote{In our UGRec, we also make a try to model co-occurrence relations as pairs of directed relation triplets just like the strategy used in CFKG and KTUP, but does not perform well.} The consistent improvement of UGRec demonstrates the effective design of our model. Comparing the performance between KTUP and CFKG, KTUP yields better performance than CFKG on \textit{CDs \& Vinyls} and \textit{Movies \& TVs}. This is mainly attributed to the mutual enhancement of optimizing both tasks of recommendation and KG completion in a unified manner in KTUP. However, KTUP underperforms CFKG on the datasets of \textit{Games}. The reason might be that the numbers of directed and undirected relations are very unbalanced and there are less directed relations in this dataset,  the TransE used in CFKG is more suitable for such situations. In contrast, our UGRec model can achieve the best  performance under different conditions, further demonstrating the advantages of our model.

\begin{figure*}[htbp]
\subfigure[Games]{\includegraphics[width=.3\linewidth]{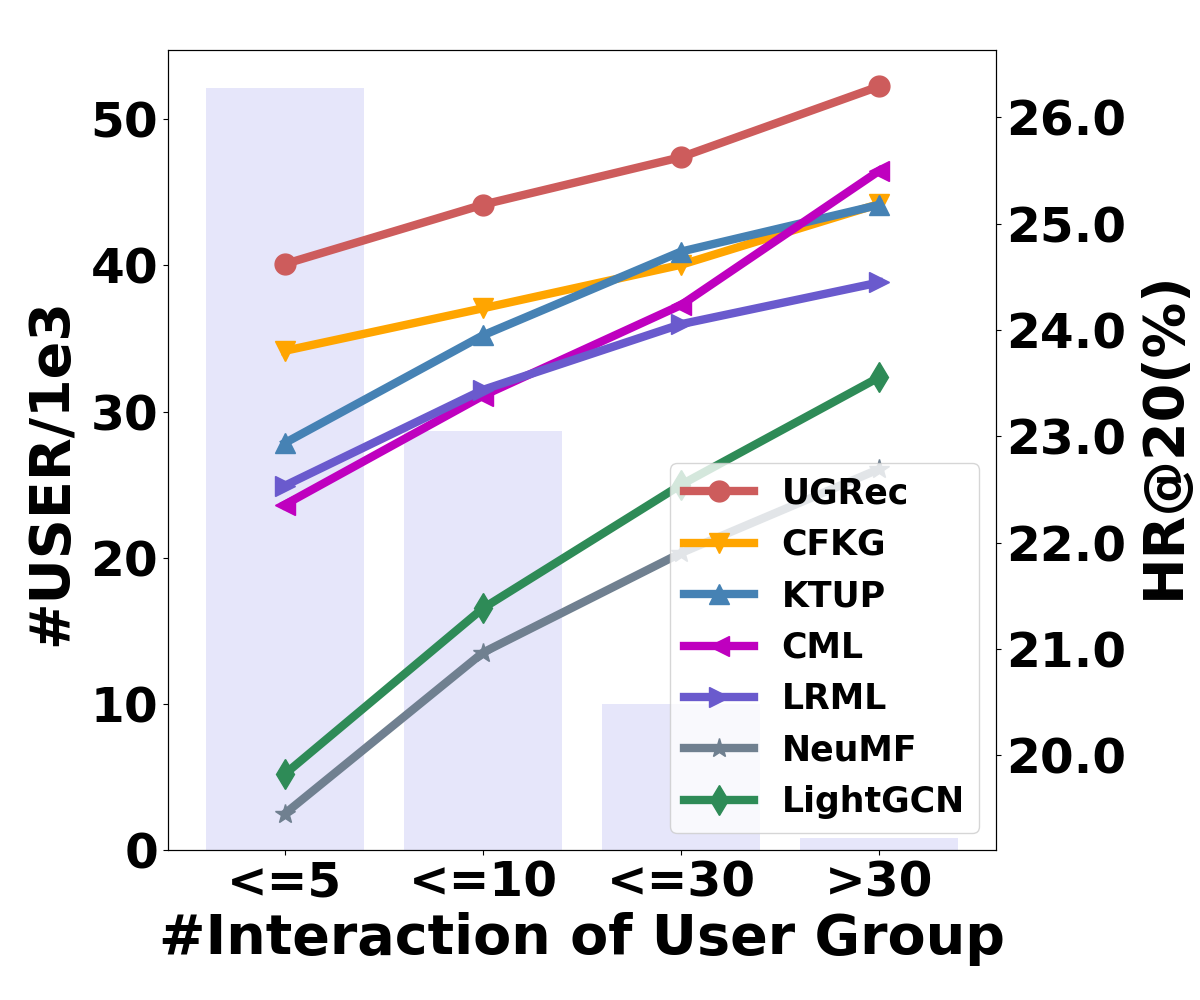}}
\subfigure[CDs \& Vinyls]{\includegraphics[width=.3\linewidth]{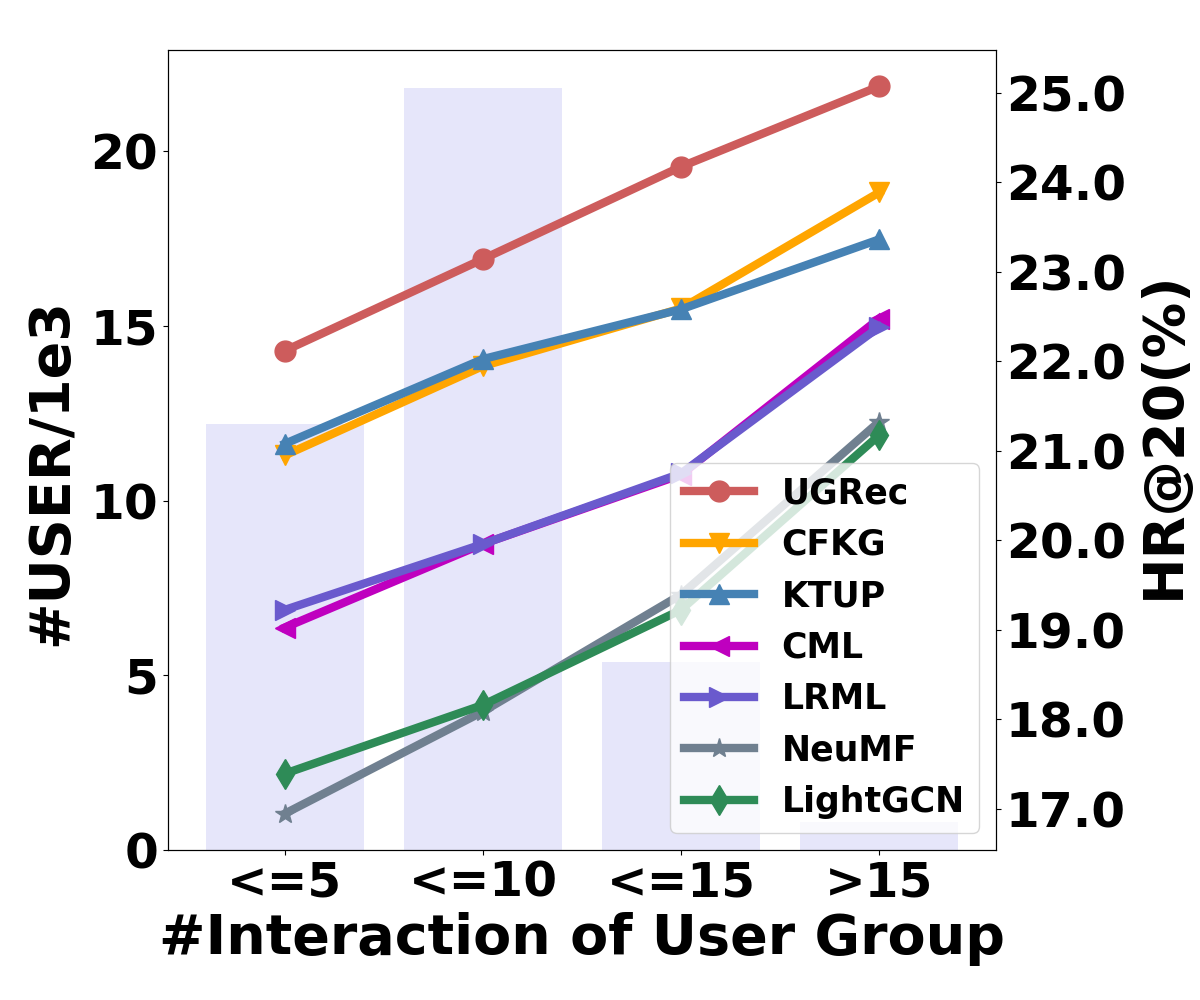}}
\subfigure[Movies \& TVs]{\includegraphics[width=.3\linewidth]{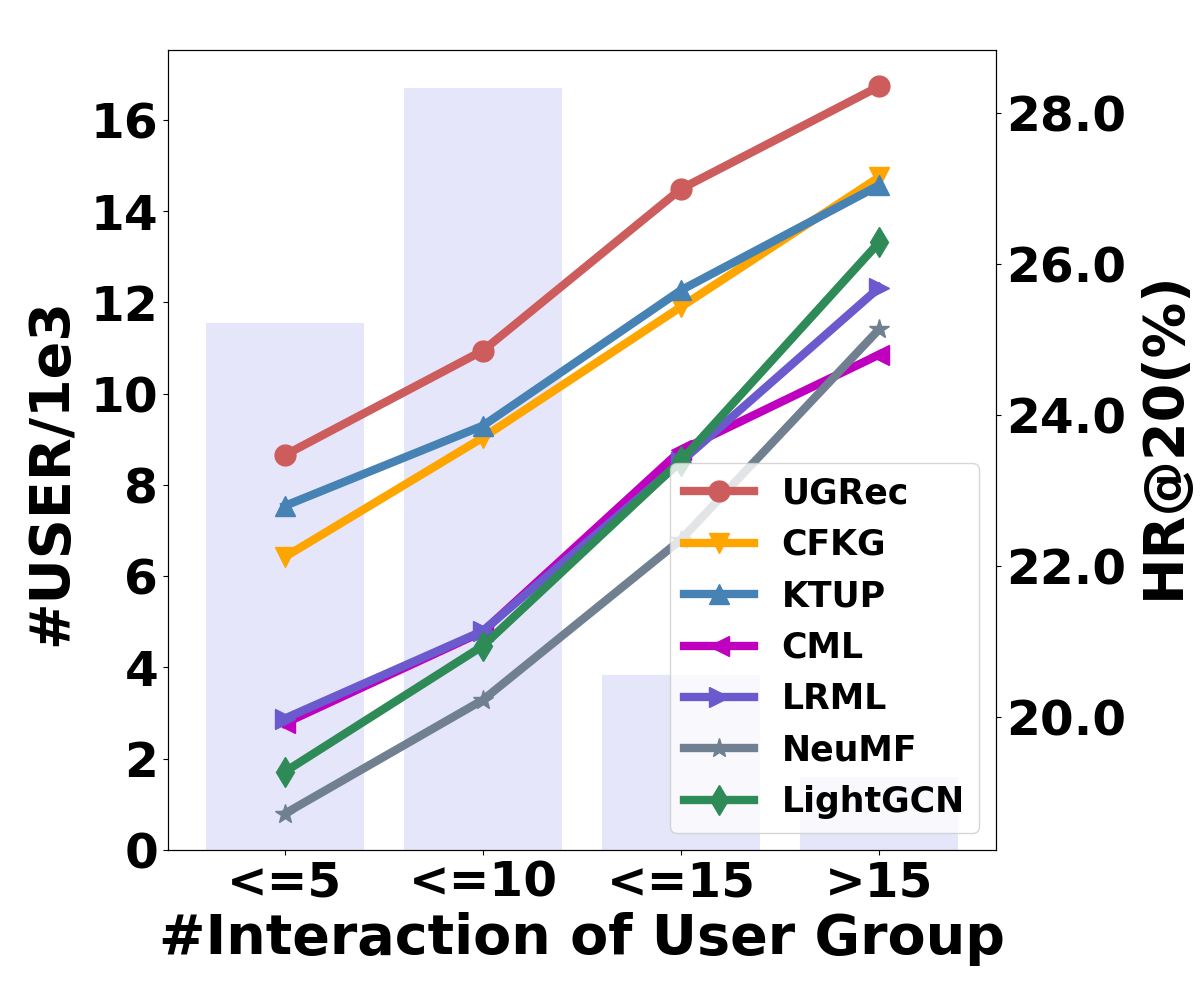}}
\vspace{-15pt}
\caption{Performance comparison over user groups of different  sparsity levels on three datasets. The background histograms indicate the number of users in each group; meanwhile, the lines show the performance \emph{w.r.t.} HR@20.}
\label{fig:sparsity}
\end{figure*}

\subsubsection{Effects of Co-occurrence Data (RQ2)} In this section, we study the effects of integrating the co-occurrence data with knowledge information in recommendation. Specifically, for each dataset, we randomly sample co-occurrence relation triplets with a ratio of $\{0,0.2,0.4,0.6, 0.8,1.0\}$, and training three models UGRec, CFKG and KTUP based on the newly constructed datasets. $ratio=0$ indicates no co-occurrence information has been used in training, and $ratio=1$ means that all the directed and undirected co-occurrence relations are used. Figure \ref{fig:co} reports the results of the three methods on each dataset.  From the results, it can be observed that our model consistently outperforms the other two KG-based models by a large margin across different ratios of co-occurrence data over all three datasets. It is worth mentioning that even only the directed knowledge relations are used (i.e., $ratio = 0$), our UGRec model can also beat the other two models, which indicates the superiority of our directed relation modeling strategy with the designed head-tail relation-aware attention mechanism. More co-occurrence relation triplets used in the training give better performance of all the models. This validates the positive effects of integrating co-occurrence data with the knowledge information for recommendation. Moreover, our model clearly enjoys a higher increasing rate, demonstrating our model is more effective on exploiting both directed and undirected relations by modeling them separately.

\subsubsection{Effects of Data sparsity (RQ3)}
The motivation of combining co-occurrence data with knowledge information is to further enhance the performance on dealing with data sparsity. In this section, we study the capability of our UGRec model on alleviating this issue and classify users into four groups according to the number  of interactions in the training set: $\leq 5$, $\leq 10$, $\leq 15$ and $> 15$ ( and $\leq 5$, $\leq 10$, $\leq 30$ and $> 30$ for the \textit{Games} dataset). Figure \ref{fig:sparsity} reports the performance \emph{w.r.t.}  HR@20 on different user groups on the three datasets. As we can see,  most users have less than 10 interactions over all the three datasets, which further validates the common sparsity problem faced by real systems. From the results, we can have the following observations. Firstly, with the increasing number of interactions, the performance of all models consistently becomes better, especially for the ones only exploiting interaction information, such as CML, LightGCN and NeuMF. Besides, the models which exploit side information obtain better performance than the ones without exploiting the information; and a higher improvement can be observed for the user groups with sparser interactions (e.g., $\leq 5$). This evidence
strongly justifies the effectiveness of side information on alleviating the sparsity issue. Furthermore, our UGRec model consistently outperforms all the baselines by a large margin over user groups of different sparsity levels. This indicates that our model exploits the knowledge and co-occurrence information more effectively than CFKG and KTUP on alleviating the sparsity problem, demonstrating the potential of our strategy on modeling the directed and undirected relations for recommendation.

\subsubsection{Ablation Study (RQ4)} To investigate the contributions of different components on the final performance, including undirected/directed relations and the attention modules, we conduct ablation studies to compare the performance of different variants of our UGRec model on the three datasets.\footnote{To save space, we just report the performance of several most typical variants.} The variants include:
\begin{itemize}[leftmargin=*]
\item \textbf{o/dc}: It only keeps the user-item interaction information in UGRec. In other words, it removes all other directed (except the \emph{interaction} relation) and undirected relation from UGRec.
\item \textbf{o/c}: It removes all the undirected co-occurrence relations from UGRec.
\item \textbf{o/d}: It removes all the directed relations (except the \emph{interaction} relation) from UGRec.
\item \textbf{o/att}: This model removes all the head-tail relation-aware attention module from UGRec. Notice that the above models still adopt the corresponding attention modules for the kept relations.
\end{itemize}
\begin{table}
  \caption{Performance of the ablation studies. The best results are highlighted in bold. HR and NDCG indicates HR@20(\%) and NDCG@20(\%) in this table.}
  \label{tab:ablation}
  \setlength{\tabcolsep}{1.6mm}
  \begin{tabular}{c|cc|cc|cc}
    \hline
    \multirow{2}{*}{}&\multicolumn{2}{c}{Games}&\multicolumn{2}{|c}{CDs \& Vinyls}&\multicolumn{2}{|c}{Movies \& TVs}\\
    \cline{2-7}
    &HR&NDCG&HR&NDCG&HR&NDCG\\
    \hline\hline
    \textbf{o/dc}&23.02&13.07&20.37&11.23&21.93&13.13\\
    \textbf{o/c}&24.27&13.64&22.47&12.25&23.94&14.10\\
    \textbf{o/d}&24.21&13.59&22.10&12.06&23.77&14.02\\
    \textbf{o/att}&24.35&13.68&22.83&12.43&24.44&14.44\\
    \textbf{UGRec}&\textbf{24.92}&\textbf{13.99}&\textbf{23.01}&\textbf{12.57}&\textbf{24.79}&\textbf{14.60}\\
    \hline
  \end{tabular}
\end{table}
Table~\ref{tab:ablation} presents the performance of all variants and our UGRec model on the three datasets. By comparing different pairs of variants, we can analyze the impact of different components on the performance. Specifically, we can have following observations:
\begin{itemize}[leftmargin=*]
\item Firstly, both o/c and o/d outperform o/dc, demonstrating the value of side information for recommendation. Although the undirected co-occurrence relations  are much richer than the directed knowledge relations, o/c still yields better performance than o/d. This is acceptable because the co-occurrence information provides auxiliary collaborative signals for the modeling as the user-item interactions. In contrast, the knowledge relations bring extra semantic information for the items, which contributes direct item features for the model.
\item In addition, although removing the attention modules, o/att can still achieve better performance than both o/d and o/c, which enjoy the benefit from the attention module. The superior performance of o/att over o/d and o/c well validate the effectiveness of combining both knowledge and co-occurrence information.
\item The much better performance of UGRec over that of o/att shows the positive effects of our designed attention method on the performance. This also indicates the importance of modeling the fine-grained preference in relation modeling. Another interesting observation is that o/dc can achieve better performance than the CML model (shown in Table~\ref{tab:result}), which further verifies the effectiveness of the attention mechanism. Note that o/dc also only exploits the interaction data and can be regarded as a variant collaborative metric learning approach with our designed attention mechanism.
\end{itemize} 

\section{RELATED WORK}
In this section, we make a brief review of the methods which exploit knowledge-graph (KG) and item-item co-occurrence information in recommendation.

\textbf{KG-based.} Knowledge-graph based recommendation has attracted lots of attention in recent years. Existing KG based recommender systems mainly apply KGs in two ways: the embedding-based approach and path-based approach. Methods of the former one often apply the KG embedding techniques to learn the user/item embeddings~\cite{a20,a28, a41}. For example, CKE~\cite{a41} applies TransR to learn the semantic embedding from KG and integrates the learned embedding with the latent factor learned from MF for recommendation. DKE~\cite{dkn} combines the knowledge-aware embedding with the word embedding of each word within the news content for news recommendation. Zhang et al.~\cite{a20} adopted the TransE to model both the directed knowledge and undirected co-occurrence  relations for entity embedding learning. Cao et al.~\cite{a28} employed the TransH to learn different representations of entities on different relations and proposed a KTUP method which jointly models the recommendation task and KG completion for performance enhancement. Recently, with the application of GCN techniques in recommendation, the GCN is also used in KG-based recommendation to learn user and item embeddings~\cite{kgat,wu2020joint}. Wang et al.~\cite{kgat} reported a knowledge graph attention network (KGAT) to model the high-order connectivities in KG with an end-to-end fashion.  Wu et al.~\cite{wu2020joint} proposed an adaptive GCN model for the joint task of item recommendation and attribute inference.

Most path-based methods exploit the semantic entities in different meta-paths as graph regularization to facilitate the user and item representation learning. For example, Hete-MF~\cite{yu2013collaborative} uses the item-item similarity learned from different meta-paths as a regularization to the MF method. Hete-CF~\cite{luo2014hete} extends this method to consider item-item, user-user, and item-user similarities learned from different meta-paths as regularization information. HeteRec~\cite{yu2013rec} leverages the meta-path similarity to enrich the user-item interaction for better performance. There are also methods which explicitly learn the path embedding to directly model the user-item relations for recommendation, such as MCRec~\cite{hu2018leve} and KPRN~\cite{wang2019aaai}. Besides, some researchers attempt to combine the two approaches for better performance. For instance, RippleNet~\cite{ripplenet} combines the advantages of the embedding-based and path-based methods and achieves a remarkable performance improvement.

A more comprehensive review for KG-based recommendation can refer to~\cite{guo2020}. The proposed UGRec method in this paper falls into the first approach, and it adopts different strategies to model both directed knowledge relations and the undirected item-item co-occurrence relations for entity embedding learning in a unified graph-based model.

\textbf{Item-item co-occurrence based.} The co-occurrence information of items has been widely used for the substitute recommendation task~\cite{a49,chen2020try}, however, it is not our target task in this work. We focus on the methods for the traditional recommendation task (e.g., top-$n$ recommendation). Park et al.~\cite{a25}  proposed a matrix co-factorization method which jointly factorizes user ratings data and co-view product information. Liang et al.~\cite{a57} used the item-item co-occurrence matrix to regularize the item representations learned from MF. In CFKG~\cite{a20}, the item co-occurrence relations are modeled as the same as the directed knowledge relations by using TransE. Xin et al.~\cite{a9} pointed out that the use of TransE for undirected relations may lead to a trivial solution, and they apply the DistMulti strategy to model the co-occurrence relations in their RCF model. Note that the RCF model only exploits the item-item relations without the directed knowledge relations. Similar to CFKG, our UGRec model also exploits both the directed knowledge relations and undirected item-item co-occurrence relations. Different from CFKG, UGRec uses different methods to model the two types of relations. In particular, the two entities of an undirected relation are projected to a unique hyperplane to minimize their distance.

\section{CONCLUSION}
In this work, we highlight the potential of integrating both knowledge and item-item co-occurrence information for enhancing the recommendation performance. As a concrete contribution, we  present a unified graph-based recommendation model (UGRec), which simultaneously models the directed knowledge relations and undirected co-occurrence relations in an end-to-end manner. In particular, our UGRec model adopts two different strategies to model the directed and undirected relations for accurate user preference modeling. In addition,  a head-tail relation-aware attention mechanism is introduced into the model for  fine-grained relation modeling. Extensive experiments have been conducted on three real-world datasets, and show that our UGRec model outperforms several competitive baselines, demonstrating the advantage of exploiting both types of information for recommendation. The superior performance over two recent KG-based recommendation models validates the effective design of UGRec on separately modeling the directed and undirected relations.

\section{ACKNOWLEDGMENTS}
This work is supported by the National Natural Science Foundation of China, No.:61902223; the Innovation Teams in Colleges and Universities in Jinan, No.:2020GXRC040; Young creative team in
universities of Shandong Province, No.:2020KJN012; New AI project towards the integration of education and industry in QLUT, No.:2020KJC-JC01; Youth Program of National Natural Science Foundation of China, No.:72004127.

\newpage
\bibliographystyle{ACM-Reference-Format}
\bibliography{ugrec-base}
\appendix
\end{document}